\documentclass[11pt, lettersize]{article}

\usepackage{epsfig}
\usepackage{amssymb}
\usepackage{amsmath}
\usepackage{times}

\usepackage{comment}
\usepackage{setspace}

\usepackage{algorithm}
\usepackage{algorithmic}
\renewcommand{\algorithmicrequire}{Input:}
\renewcommand{\algorithmicensure}{Output:}

\newtheorem{theorem}{\bf Theorem}

\newtheorem{definition}{\bf Definition}

\begin{document}

\pagestyle{plain}
\pagenumbering{arabic}
\setcounter{page}{1}

\begin{center}
{\Large \bf Efficient 2-Step Protocol and Its Discriminative Feature Selections \\ in Secure Similar Document Detection}

Sang-Pil Kim$^{\dagger}$, Myeong-Sun Gil$^{\dagger}$, Yang-Sae Moon$^{\dagger}$, and Hee-Sun Won$^{\ddagger}$

$^{\dagger}$Department of Computer Science, Kangwon National University \\%
\vspace*{-0.15cm}
1 Kangwondaehak-gil, Chuncheon-si, Gangwon 200-701, Republic of Korea

$^{\ddagger}$Electronics and Telecommunications Research Institute\\%
\vspace*{-0.15cm}
218 Gajeong-ro, Yuseong-gu, Daejeon 305-701, Republic of Korea

E-mail: \{spkim, gils, ysmoon\}@kangwon.ac.kr, hswon@etri.re.kr
\end{center}


\begin{abstract}

Secure similar document detection\,(SSDD) identifies similar documents of two parties while each party does not disclose its own {\em sensitive} documents to another party. In this paper, we propose an efficient 2-step protocol that exploits a feature selection as the lower-dimensional transformation and presents discriminative feature selections to maximize the performance of the protocol. For this, we first analyze that the existing 1-step protocol causes serious computation and communication overhead for high dimensional document vectors. To alleviate the overhead, we next present the feature selection-based 2-step protocol and formally prove its correctness. The proposed 2-step protocol works as follows: (1) in the {\it filtering\/} step, it uses low dimensional vectors obtained by the feature selection to filter out non-similar documents; (2) in the {\it post-processing\/} step, it identifies similar documents only from the non-filtered documents by using the 1-step protocol. As the feature selection, we first consider the simplest one, random projection\,(RP), and propose its 2-step solution {\sf SSDD-RP}. We then present two discriminative feature selections and their solutions: {\sf SSDD-LF}\,(local frequency) which selects a few dimensions locally frequent in the current querying vector and {\sf SSDD-GF}\,(global frequency) which selects ones globally frequent in the set of all document vectors. We finally propose a hybrid one, {\sf SSDD-HF}\,(hybrid frequency), that takes advantage of both {\sf SSDD-LF} and {\sf SSDD-GF}. We empirically show that the proposed 2-step protocol outperforms the 1-step protocol by three or four orders of magnitude. \\

\noindent%
\end{abstract}

\section{Introduction} \label{sec:sec1}
\begin{figure}[hbt]
  \centering
  \psfig{file=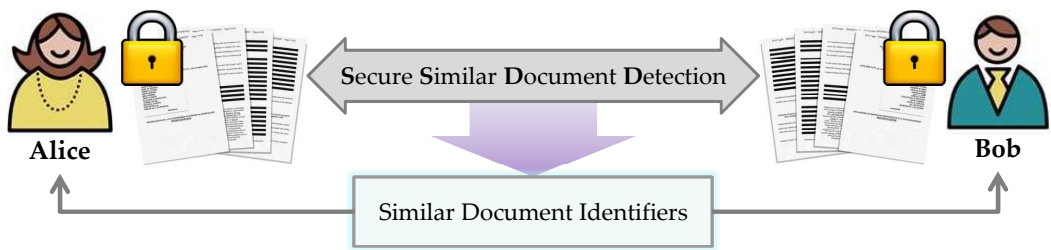,width=5.50in}\\
  \caption{Concept of secure similar document detection.}
  \label{fig:fig1}
\end{figure}

%
Similar document detection is the problem of finding similar documents of two parties, Alice and Bob, and it has been widely used in version management of files, copyright protection, and plagiarism detection\,\cite{Shi95,Soro96}. Recently, secure similar document detection\,(SSDD)\,\cite{Jia08} has been introduced to identify similar documents while preserving privacy of each party's documents as shown in Figure \ref{fig:fig1}. That is, SSDD finds similar document pairs whose cosine similarity\,\cite{Han06,Sta07} exceeds the given tolerance while not disclosing document vectors to each other party. SSDD is a typical example of privacy-preserving data mining\,(PPDM)\,\cite{Agg07,Agr00,Lin02}, and has the following applications\,\cite{Jia08}. First, in two or more conferences that are not allowing double submissions, SSDD finds the double-submitted papers while not disclosing the papers to each other conference. Second, in the insurance fraud detection system, SSDD searches similar accident cases of two or more insurance companies while not providing sensitive and private cases to each other company.

Jiang~et~al.\,\cite{Jia08} have proposed a novel solution for SSDD by exploiting secure multiparty computation\,(SMC)\,\cite{Cli02,Pin02} in a semi-honest model. Their solution has preserved privacy of two parties by using the secure scalar product in computing cosine similarity between document vectors. As the secure scalar product, they have suggested random matrix and homomorphic encryption methods\,\cite{Goe04,Vai02}. In this paper, we use the random matrix method as a base protocol, and we call it {\sf SSDD-Base}. However, {\sf SSDD-Base} has a critical problem of incurring severe computation and communication overhead. Let Alice's and Bob's document sets be $\mathbb{U}$ and $\mathbb{V}$, respectively, then {\sf SSDD-Base} requires $|\mathbb{U}||\mathbb{V}|$ secure scalar products. In many cases, the dimension $n$ of document vectors reaches tens of thousands or even hundreds of thousands, and {\sf SSDD-Base} incurs a very high complexity of $O(n|\mathbb{U}||\mathbb{V}|)$, which is not practical to support a large volume of document databases. In particular, if there are many parties or frequent changes in document databases, the overhead becomes much more critical.

To alleviate the computation and communication overhead of {\sf SSDD-Base}, in this paper we present a 2-step protocol that exploits the feature selection of lower-dimensional transformation. The feature selection transforms high dimensional document vectors to low dimensional feature vectors, and in general it selects tens to hundreds dimensions from thousands to tens of thousands dimensions. We call the feature selection {\it FS} in short. Representative FS includes RP\,(random projection)\,\cite{Bin01}, DF\,(document frequency)\,\cite{Tang05}, and LDA\,(linear discriminant analysis)\,\cite{Cai08}. In this paper, we use RP and DF since they are known as simple but efficient feature selections\,\cite{Tang05}. To devise a 2-step protocol, we need to find an upper bound of cosine similarity for the filtering process. Thus, we first present an upper bound of FS and formally prove its correctness. Using the upper bound property of FS, we then propose a generic 2-step protocol, called {\sf SSDD-FS}. The proposed {\sf SSDD-FS} works as follows: in the first {\it filtering\/} step, it converts $n$-dimensional vectors to $f(\ll n)$-dimensional vectors and applies the secure protocol to $f$-dimensional vectors to filter out non-similar $n$-dimensional vectors; in the second {\it post-processing\/} step, it applies the base protocol {\sf SSDD-Base} to the non-filtered $n$-dimensional vectors. In the filtering step, {\sf SSDD-FS} prunes many non-similar {\it high\/} dimensional vectors by comparing {\it low\/} dimensional vectors with relatively less complexity of $O(f|\mathbb{U}||\mathbb{V}|)$, and thus, it significantly improves the performance compared with {\sf SSDD-Base}.

To make {\sf SSDD-FS} be efficient, FS should be highly discriminative, i.e., FS should filter out as many high dimensional vectors as possible if they are non-similar. In this paper, we analyze SSDD protocols in detail and propose four different techniques as the discriminative implementation of FS. We can think RP first as an easiest way of implementing FS. RP randomly selects $f$ dimensions from $n$ dimensions. RP is easy, but its filtering effect will be very low due to the randomness. To solve the problem of RP, we exploit DF that selects feature dimensions based on frequencies in all document vectors. In particular, by referring the concept of DF, we present three variants of DF, called LF\,(local frequency), GF\,(global frequency), and HF\,(hybrid frequency). First, LF considers term frequencies of Alice's current querying vector\,(we call it the {\it current vector}), and it selects dimensions whose frequencies higher than the others in the current vector. LF focuses on the {\it locality\/}, which means that considering the current vector only might be enough to decrease the upper bound of cosine similarity. Second, GF means DF itself, that is, GF counts the number of documents containing each term\,(dimension), constructs a frequency vector from those counts\,(we call it the {\it whole vector}), and selects high frequency dimensions from the whole vector. GF focuses on the {\it globality} since it considers all the document vectors. To implement GF, however, we need to make a secure protocol for obtaining the whole vector from both Alice's and Bob's document sets. For this, we propose a secure protocol {\sf SecureDF} as a secure implementation of DF. Third, HF takes advantage of both locality of LF and globality of GF. HF computes a {\it difference vector\/} between current and whole vectors and selects high-valued dimensions from the difference vector. This is because HF tries to maximize the value difference between Alice's and Bob's vectors for each selected dimension and eventually decrease the upper bound of cosine similarity. Table \ref{tbl:tbl1} summarizes these four feature selections and their corresponding SSDD protocols, {\sf SSDD-RP}, {\sf SSDD-LF}, {\sf SSDD-GF}, and {\sf SSDD-HF}, to be proposed in Section \ref{sec:sec4}.

\begin{table}
\begin{center}
\caption{Feature selection methods to be used for {\sf SSDD-FS}.} \label{tbl:tbl1}%
\begin{tabular}{|c|l|} \hline
Method        & \multicolumn{1}{|c|}{Description} \\ \hline \hline %
{\sf SSDD-RP} & Randomly select $f$ dimensions from an $n$-dimensional vector. \\ \hline %
{\sf SSDD-LF} & Select highly frequent $f$ dimensions from Alice's $n$-dimensional current vector. \\ \hline %
{\sf SSDD-GF} & Select highly frequent $f$ dimensions from the $n$-dimensional whole vector. \\ \hline %
{\sf SSDD-RP} & Select high-valued $f$ dimensions from the $n$-dimensional difference vector between \\
              & current and whole vectors. \\ \hline %
\end{tabular}
\end{center}
\end{table}

In this paper, we empirically evaluate the base protocol, {\sf SSDD-Base}, and our four {\sf SSDD-FS} protocols\,({\sf SSDD-RP}, {\sf SSDD-LF}, {\sf SSDD-GF}, {\sf SSDD-HF}) using various data sets. Experimental results show that the {\sf SSDD-FS} protocols significantly outperform {\sf SSDD-Base}. This means that the proposed 2-step protocols effectively prune a large number of non-similar sequences early in the filtering step. In particular, {\sf SSDD-HF} that takes advantage of both locality of {\sf SSDD-LF} and globality of {\sf SSDD-GF} shows the best performance. Compared with {\sf SSDD-Base}, {\sf SSDD-HF} reduces the execution time of SSDD by three or four orders of magnitude.

The rest of this paper is organized as follows. Section~\ref{sec:sec2} explains related work and background of the research. Section~\ref{sec:sec3} presents the FS-based 2-step protocol, {\sf SSDD-FS}, and proves its correctness. Section~\ref{sec:sec4} introduces four novel feature selections, RP, LF, GF, and HF, and it proposes their corresponding secure protocols. Section~\ref{sec:sec5} explains experimental results on various data sets. We finally summarize and conclude the paper in Section~\ref{sec:sec6}.

\section{Related Work and Background} \label{sec:sec2}
%
We use cosine similarity as the basic operation of similar document detection. The cosine similarity of two $n$-dimensional vectors $\overrightarrow{U} = \{u_1, \ldots, u_n\}$ and $\overrightarrow{V} = \{v_1, \ldots, v_n\}$ is computed as $\cos(\overrightarrow{U},\overrightarrow{V}) = \frac{\overrightarrow{U} \cdot \overrightarrow{V}}{\left\|U\right\| \cdot \left\|V\right\|}$, where $\overrightarrow{U} \cdot \overrightarrow{V}$ is the scalar product of $\overrightarrow{U}$ and $\overrightarrow{V}$, that is, $\overrightarrow{U} \cdot \overrightarrow{V} = \sum_{i=1}^{n} u_i v_i$. If we can compute $\overrightarrow{U} \cdot \overrightarrow{V}$ securely in two parties, we can also compute $\cos(\overrightarrow{U},\overrightarrow{V})$ securely. There are two representative methods for the secure scalar product\,\cite{Jia08}. The first one is the random matrix method\,\cite{Vai02}, where two parties share the same random matrix and compute the scalar product securely using the matrix. The second one is the homomorphic encryption method\,\cite{Goe04}, where two parties use the homomorphic probability key system for the secure computation of scalar products. In this paper, we use the random matrix method since it is more efficient than the homomorphic encryption one, but we can also instead use the homomorphic encryption method for the protocols to be discussed later. Without loss of generality, we assume that vectors $\overrightarrow{U}$ and $\overrightarrow{V}$ are normalized to size $1$. That is, $\left\|U\right\| = \left\|V\right\| = 1$, and thus, simply $\cos(\overrightarrow{U},\overrightarrow{V}) = \overrightarrow{U} \cdot \overrightarrow{V}$.

\begin{figure}[hbt]
  \centering
  \psfig{file=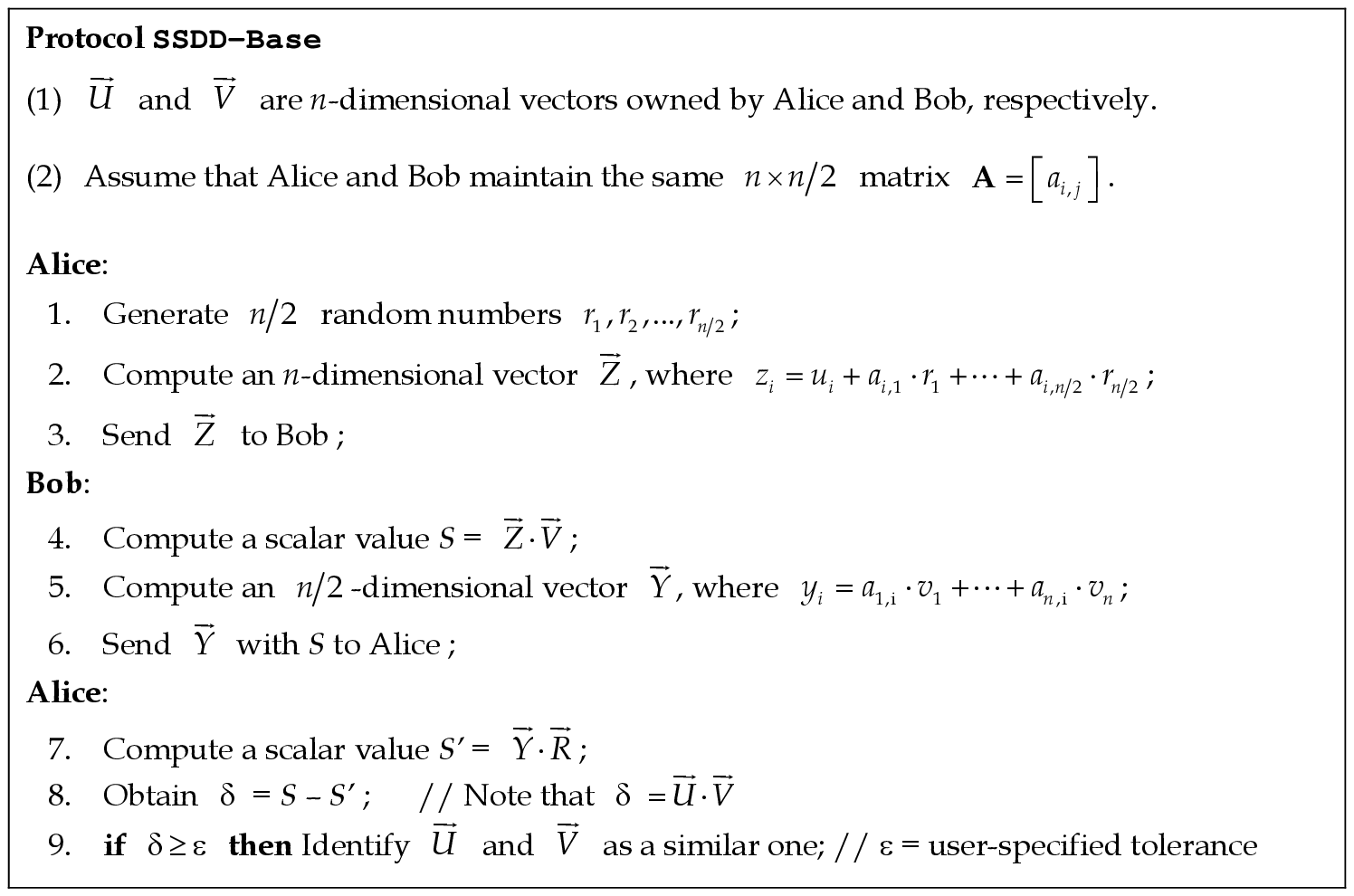,width=6.00in}\\
  \caption{Protocol of {\sf SSDD-Base}.}
  \label{fig:fig2}
\end{figure}

Figure~\ref{fig:fig2} shows the protocol of {\sf SSDD-Base}, the recent solution of SSDD by Jiang~et~al.\cite{Jia08}. {\sf SSDD-Base} uses the random matrix method\,\cite{Vai02} for secure scalar products, where Alice and Bob share the same matrix $\mathbf{A}$ and securely determine whether two vectors $\overrightarrow{U}$ and $\overrightarrow{V}$ are similar or not. For the correctness and detailed explanation on {\bf Protocol} {\sf SSDD-Base}, readers are referred to \cite{Jia08}. In SSDD, we perform {\sf SSDD-Base} for each pair of document vectors. More formally, if $\mathbb{U}$ and $\mathbb{V}$ are sets of document vectors owned by Alice and Bob, respectively, we perform {\sf SSDD-Base} for each pair $(\overrightarrow{U},\overrightarrow{V})$, where $\overrightarrow{U} \in \mathbb{U}$ and $\overrightarrow{V} \in \mathbb{V}$. As we mentioned in Section~\ref{sec:sec1}, however, {\sf SSDD-Base} incurs the severe computation and communication overhead of $O(n\left\|U\right\|\left\|V\right\|)$, which will be much serious if there are several parties, or a large number of documents are changed dynamically. To alleviate this critical overhead, in this paper we discuss the 2-step solution for SSDD.

In text mining and time-series mining, many lower-dimensional transformations have been proposed to solve the dimensionality curse problem\,\cite{Ber98,Moon10b} of high dimensional vectors. We can classify lower-dimensional transformations into feature extractions and feature selections\,\cite{Seb02,Tang05}. First, the feature extraction {\it creates\/} a few {\it new\/} features from an original high dimensional vector. Representative examples of feature extractions include LSI\,(latent semantic indexing)\,\cite{Dee90,Yang97}, LPI\,(locality preserving indexing)\,\cite{Cai05}, DFT\,(discrete Fourier transform)\,\cite{Fal94,Loh10,Moon14}, DWT\,(discrete Wavelet transform)\,\cite{Chan03,Moon02}, and
PAA\,(piecewise aggregate approximation)\,\cite{Han11,Yi00}. In contrary, the feature selection {\it selects\/} a few {\it discriminative\/} features from an original (or transformed) high dimensional vectors. Representative examples of feature selections include RP, DF, LDA, and PCA\,(principal component analysis)\,\cite{Bin01,Cai08,Tang05}. In this paper, we use RP and DF with appropriate variations. This is because RP and DF are much simpler than other transformations, and accordingly, they are easily applied to SSDD with low complexity; on the other hand, LSI, LPI, LDA, and PCA may provide very accurate feature vectors, but they are too complex to be applied to SSDD. For the detailed explanation on lower-dimensional transformations for text mining, readers are referred to \cite{Seb02,Tang05,Yang97}.

There have been many efforts on PPDM\,\cite{Ber08}. PPDM solutions can be classified into four categories: data perturbation, $k$-anonymization, distributed privacy preservation, and privacy preservation of mining results\,\cite{Moon10a}. SSDD can be regarded as an application of distributed privacy privation. For the detailed explanation on problems and solutions of data perturbation and $k$-anonymization, readers are referred to survey papers\,\cite{Agg07,Ber08}.

\section{Feature Selection-based Secure 2-Step Protocol} \label{sec:sec3}
%
In this paper, we use FS, feature selection, for the secure 2-step protocol. To transform an $n$-dimensional vector to an $f$-dimensional vector, FS chooses randomly or highly frequent $f$ dimensions from $n$ dimensions, and thus, its transformation process is very simple. In this section, we first assume that FS can select $f$ dimensions from $n$ dimensions in a secure manner, and we then propose the secure 2-step protocol of SSDD by using the secure FS.

To use a lower-dimensional transformation $F$ for SSDD, we need to find an upper bound function $upper(\overrightarrow{U^F},\overrightarrow{V^F})$ that satisfies Eq.~(\ref{eq:eq1}), where $\overrightarrow{U^F}$ and $\overrightarrow{V^F}$ are $f$-dimensional feature vectors selected from $n$-dimensional vectors, $\overrightarrow{U}$ and $\overrightarrow{V}$, respectively, by the transformation $F$. In Eq.~(\ref{eq:eq1}), $\overrightarrow{U} = \{u_1, \ldots, u_n\}$, $\overrightarrow{V} = \{v_1, \ldots, v_n\}$, $\overrightarrow{U^F} = \{u_1^F, \ldots, u_f^F\}$, and $\overrightarrow{V^F} = \{v_1^F, \ldots, v_f^F\}$.
\begin{equation}
\label{eq:eq1}
\cos\!\left(\overrightarrow{U},\overrightarrow{V}\right)
  \leq upper\!\left(\overrightarrow{U^F},\overrightarrow{V^F}\right).
\end{equation}
The reason why the transformation $F$ should satisfy Eq.~(\ref{eq:eq1}) is that SSDD of using $F$ should not incur any false dismissal, and this is known as Parseval's theorem\,(the lower bound property of Euclidean distances) in time-series matching\,\cite{Fal94,Moon02,Moon10b}. To obtain an upper bound of the lower-dimensional transformation $F$, we first define an upper bound of $F$ as follows.
\begin{definition}
\label{def:def1} {\rm
If a lower-dimensional transformation $F$ transforms $n$-dimensional vectors, $\overrightarrow{U}$ and $\overrightarrow{V}$, to $f$-dimensional vectors, $\overrightarrow{U^F}$ and $\overrightarrow{V^F}$, respectively, we define an {\it upper bound function\/} of $F$, denoted by $upper(\overrightarrow{U^F},\overrightarrow{V^F})$, as Eq.~(\ref{eq:eq2}).}
\begin{equation}
\label{eq:eq2}
upper\!\left(\overrightarrow{U^F},\overrightarrow{V^F}\right)
  = 1 - \frac{D^2\!\left(\overrightarrow{U^F},\overrightarrow{V^F}\right)}{2},
\end{equation}
{\rm where $D^2(\overrightarrow{U^F},\overrightarrow{V^F})$ is the squared Euclidean distance between $\overrightarrow{U^F}$ and $\overrightarrow{V^F}$, i.e., $D^2(\overrightarrow{U^F},\overrightarrow{V^F}) = \sum_{i=1}^{f}|u_i^F - v_i^F|^2$. $\fbox{}$}
\end{definition}
In this paper, we want to use FS as a lower-dimensional transformation $F$, and thus, we formally prove that the upper bound function of FS satisfies Eq.~(\ref{eq:eq1}), the upper bound property of cosine similarity.

\begin{theorem}
\label{th:th1}
If a feature selection FS transforms $n$-dimensional vectors, $\overrightarrow{U}$ and $\overrightarrow{V}$, to $f$-dimensional vectors, $\overrightarrow{U^{FS}}$ and $\overrightarrow{V^{FS}}$, respectively, $upper(\overrightarrow{U^{FS}},\overrightarrow{V^{FS}})$ is an upper bound of $\cos(\overrightarrow{U},\overrightarrow{V})$, that is, Eq.~{\rm (\ref{eq:eq3})} holds.
\begin{equation}
\label{eq:eq3}
\cos\!\left(\overrightarrow{U},\overrightarrow{V}\right)
  \leq upper\!\left(\overrightarrow{U^{FS}},\overrightarrow{V^{FS}}\right).
\end{equation}
\end{theorem}
{\sc Proof}: First, let $\overrightarrow{U} = \{u_1, \ldots, u_n\}$, $\overrightarrow{V} = \{v_1, \ldots, v_n\}$, $\overrightarrow{U^{FS}} = \{u_1^{FS}, \ldots, u_f^{FS}\}$, and $\overrightarrow{V^{FS}} = \{v_1^{FS}, \ldots, v_f^{FS}\}$. Then, Eqs.~(\ref{eq:eq4}) and (\ref{eq:eq5}) hold for $\overrightarrow{U}$ and $\overrightarrow{V}$.
\begin{eqnarray}
D^2\!\left(\overrightarrow{U},\overrightarrow{V}\right) & = & \sum_{i=1}^{n}(u_i - v_i)^2 \nonumber \\
  & = & \sum_{i=1}^{n}u_i^2 - 2\cdot\sum_{i=1}^{n}u_iv_i + \sum_{i=1}^{n}v_i^2 \nonumber \\
  & = & \left\|\overrightarrow{U}\right\| - 2\left|\overrightarrow{U}\right|\left|\overrightarrow{V}\right|
    + \left\|\overrightarrow{V}\right\| \nonumber \\
  & = & 2 - 2\cos\left(\overrightarrow{U},\overrightarrow{V}\right). \label{eq:eq4}
\end{eqnarray}
\begin{equation}
\label{eq:eq5}
\cos\!\left(\overrightarrow{U},\overrightarrow{V}\right)
  = 1 - \frac{D^2\!\left(\overrightarrow{U},\overrightarrow{V}\right)}{2}.
\end{equation}
We note that all entry values of $\overrightarrow{U}$ and $\overrightarrow{V}$ are non-negative, and FS constructs $\overrightarrow{U^{FS}}$ and $\overrightarrow{V^{FS}}$ by choosing $f$ features from $\overrightarrow{U}$ and $\overrightarrow{V}$. Based on this property, Eq.~(\ref{eq:eq6}) holds.
\begin{equation}
\label{eq:eq6}
D^2\!\left(\overrightarrow{U},\overrightarrow{V}\right)
  = \sum_{i=1}^{n}(u_i - v_i)^2
  \geq \sum_{i=1}^{f}(u_i^{FS} - v_i^{FS})^2
  = D^2\!\left(\overrightarrow{U^{FS}},\overrightarrow{V^{FS}}\right).
\end{equation}
Finally, Eq.~(\ref{eq:eq7}) holds by Eqs.~(\ref{eq:eq5}), (\ref{eq:eq6}), and Eq.~(\ref{eq:eq2}) of Definition~\ref{def:def1}.
\begin{equation}
\label{eq:eq7}
\cos\!\left(\overrightarrow{U},\overrightarrow{V}\right)
  = 1 - \frac{D^2\!\left(\overrightarrow{U},\overrightarrow{V}\right)}{2}
  \leq 1 - \frac{D^2\!\left(\overrightarrow{U^{FS}},\overrightarrow{V^{FS}}\right)}{2}
  = upper\!\left(\overrightarrow{U^{FS}},\overrightarrow{V^{FS}}\right).
\end{equation}
Therefore, $upper(\overrightarrow{U^{FS}},\overrightarrow{V^{FS}})$ is an upper bound of $\cos(\overrightarrow{U},\overrightarrow{V})$. ~ $\fbox{}$

\begin{figure}[hbt]
  \centering
  \psfig{file=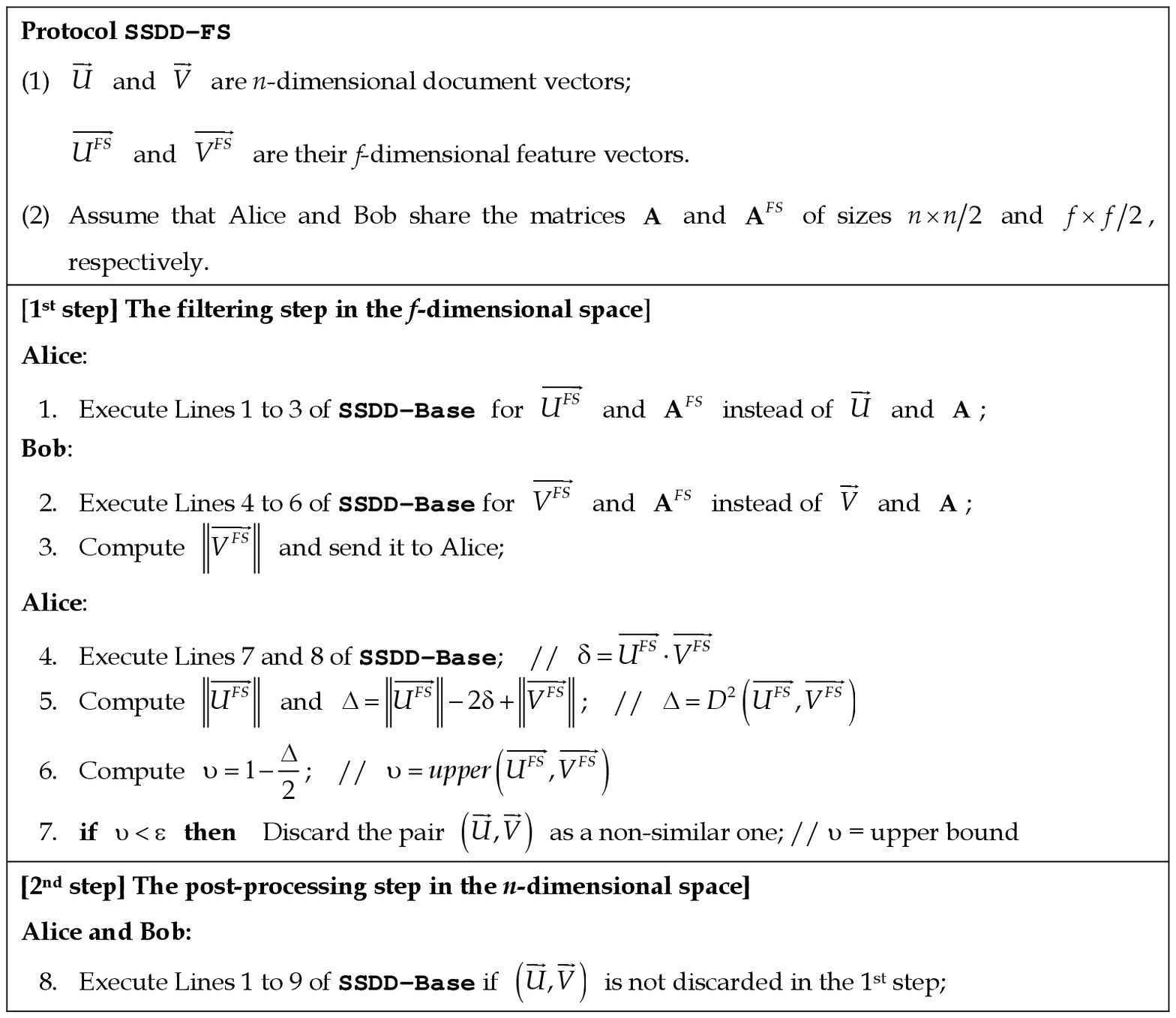,width=6.00in}\\
  \caption{Protocol of the generic 2-step solution {\sf SSDD-FS}.}
  \label{fig:fig3}
\end{figure}

By using the upper bound property of FS, we now propose a generic 2-step protocol {\sf SSDD-FS}. Figure~\ref{fig:fig3} shows {\bf Protocol} {\sf SSDD-FS}. As shown in the protocol, {\sf SSDD-FS} maintains $f$-dimensional $\overrightarrow{U^{FS}}$ and $\overrightarrow{U^{FS}}$ as well as $n$-dimensional $\overrightarrow{U}$ and $\overrightarrow{V}$ of {\sf SSDD-Base}. Also, Alice and Bob share an $f \times f/2$ matrix $\mathbf{A}\!^{FS}$ as well as an $n \times n/2$ matrix $\mathbf{A}$ of {\sf SSDD-Base}. Lines 1 to 7 of {\sf SSDD-FS} are the first step of discarding non-similar $n$-dimensional vectors in the $f$-dimensional space. First, Lines 1 to 4 securely compute the scalar product $\delta$ for $f$-dimensional vectors $\overrightarrow{U^{FS}}$ and $\overrightarrow{V^{FS}}$. Except using $f$-dimensional vectors instead of $n$-dimensional vectors, these steps are the same as those of {\sf SSDD-Base}. The only difference from {\sf SSDD-Base} is that Bob additionally sends $\left\|\overrightarrow{V^{FS}}\right\|$ to Alice in Line 3 for computing $D^2(\overrightarrow{U^{FS}},\overrightarrow{V^{FS}})$. In Line 5, Alice computes $\Delta\left(= D^2(\overrightarrow{U^{FS}},\overrightarrow{V^{FS}})\right)$ by using Eq.~(\ref{eq:eq8}).
\begin{equation}
\label{eq:eq8}
D^2\!\left(\overrightarrow{U^{FS}},\overrightarrow{V^{FS}}\right)
  = \left\|\overrightarrow{U^{FS}}\right\| - 2\cdot\overrightarrow{U^{FS}}\cdot\overrightarrow{V^{FS}}
    + \left\|\overrightarrow{V^{FS}}\right\|
  = \left\|\overrightarrow{U^{FS}}\right\| - 2\delta + \left\|\overrightarrow{V^{FS}}\right\|.
\end{equation}
After then, Alice computes an upper bound function of FS, $upper(\overrightarrow{U^{FS}},\overrightarrow{V^{FS}})$, in Line 6. In Line 7, we perform the filtering process by comparing the upper bound\,($= \upsilon$) and the given tolerance\,($= \epsilon$). If the upper bound is less than the tolerance, i.e. if $\upsilon < \epsilon$, the actual cosine similarity will also be less than the tolerance, and we don't need to compute it in the next $n$-dimensional space. That is, if $\upsilon < \epsilon$, we can skip Line 8 of the second step. Thus, Line 8 is executed only if $n$-dimensional vectors of $(\overrightarrow{U},\overrightarrow{V})$ are not filtered out by the upper bound. In Line 8, we compute the actual cosine similarity for $(\overrightarrow{U},\overrightarrow{V})$ by using {\sf SSDD-Base}.

We here note that how {\sf SSDD-FS} improves the performance compared with {\sf SSDD-Base} depends on how many $n$-dimensional vectors are discarded in the first step. This filtering effect largely depends on the discriminative power of the feature selection, i.e., efficiency of FS. In other words, if FS exploits the filtering effect largely, {\sf SSDD-FS} can reduce the computation and communication overhead from $O(n\left|U\right|\left|V\right|)$ to $O(f\left|U\right|\left|V\right|)$. Based on this observation, we need to maximize the filtering effect of FS, and this can be seen a problem of how we choose $f$ dimensions from $n$ dimensions for maximizing the discriminative power of FS. Therefore, we propose efficient FS variants and their SSDD protocols in Section~\ref{sec:sec4} and evaluate their performance in Section~\ref{sec:sec5}.

\section{Discriminative Feature Selections for the 2-Step Protocol} \label{sec:sec4}
\begin{figure}[hbt]
  \centering
  \psfig{file=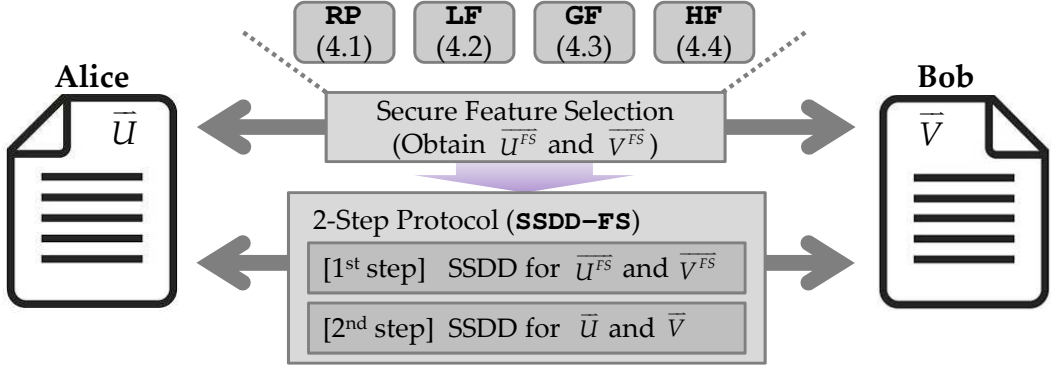,width=5.50in}\\
  \caption{Feature selections in the process of {\sf SSDD-FS}.}
  \label{fig:fig4}
\end{figure}

%
In this section, we propose four methods to implement FS of {\bf Protocol} {\sf SSDD-FS}. Figure~\ref{fig:fig4} shows the procedure of {\sf SSDD-FS} including the feature selection step. As shown in the figure, we first obtain $\overrightarrow{U^{FS}}$ and $\overrightarrow{V^{FS}}$ from $\overrightarrow{U}$ and $\overrightarrow{V}$ through the feature selection which should also be done securely. As mentioned in Section~\ref{sec:sec1}, we present RP, LF, GF, and HF as the feature selection method, and we explain how they work in detail in Sections~\ref{ssec:sec41} to \ref{ssec:sec44}.
In Figure~\ref{fig:fig4}, the secure feature selection corresponds to Line (1) of {\bf Protocol} {\sf SSDD-FS}, and the other two steps correspond to the first step\,(Lines 1 to 7) and the second step\,(Line 8), respectively.

\subsection{RP: Random Projection} \label{ssec:sec41}
%
RP is an easiest way of implementing FS, which selects $f$ dimensions randomly from $n$ dimensions. We can think two different methods in applying RP to {\sf SSDD-FS}. The first one selects $f$ dimensions dynamically for each document pair $(\overrightarrow{U},\overrightarrow{V})$; the second one first determines $f$ dimensions and then uses those pre-determined dimensions for all document pairs.

\begin{figure}[hbt]
  \centering
  \psfig{file=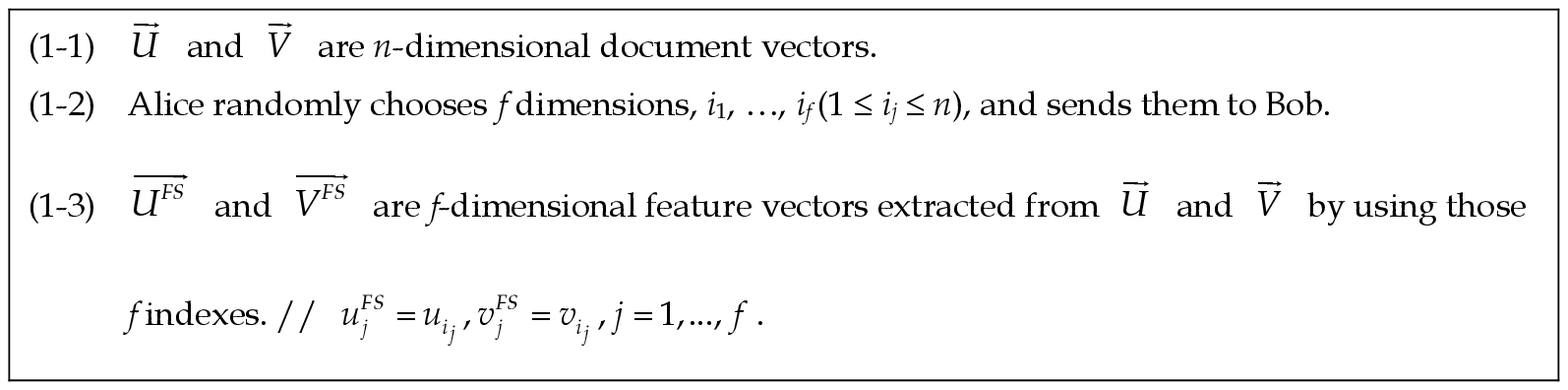,width=6.00in}\\
  \caption{Modification of Line (1) of {\sf SSDD-FS} to implement {\sf SSDD-RP}.}
  \label{fig:fig5}
\end{figure}

To use the first RP method, Alice and Bob should share $f$ indexes, $i_1, \ldots, i_f\,(1 \leq i_j \leq n, j = 1, \ldots, f)$, of randomly selected $f$ dimensions for each $(\overrightarrow{U},\overrightarrow{V})$ before starting the first step of {\sf SSDD-FS}. This sharing process can be implemented as Alice randomly selects $f$ dimensions and sends their indexes to Bob, or Alice and Bob share the same seed of the random function. That is, we can implement the first RP method by modifying Line (1) of {\bf Protocol} {\sf SSDD-FS} as Lines (1-1) to (1-3) of Figure~\ref{fig:fig5}.

The second RP method uses the same $f$ dimensions for all $(\overrightarrow{U},\overrightarrow{V})$ pairs. We can easily implement this method as Alice and Bob share the same $f$ indexes only once before starting {\sf SSDD-FS}. These first and second RP methods do not disclose any values of Alice's and Bob's document vectors, and thus, they are said to be secure. Also, these two methods have the same effect in selecting $f$ dimensions randomly. Thus, we use the second one since it is much simpler than the first one, and we call the second one {\sf SSDD-RP} by differentiating it from {\sf SSDD-FS}.

\subsection{LF: Local Frequency} \label{ssec:sec42}
%
{\sf SSDD-RP} proposed in Section~\ref{ssec:sec41} has a problem of exploiting only a little filtering effect in the first filtering step. This low filtering effect is due to that RP chooses features without any consideration of characteristics of document vectors. According to the real experiments, {\sf SSDD-RP} shows a very little improvement in SSDD performance compared with {\sf SSDD-Base}. To solve the problem of {\sf SSDD-RP} and to enlarge the filtering effect, in this paper we consider how frequent each term is in the document or document set, i.e., we use the term frequency\,(TF)\footnote{In this paper, we use TF for simplicity, but we can also use TF-IDF\,(term frequency-inverse document frequency) instead of TF. Using which frequency among TF, TF-IDF, and other feature frequencies is orthogonal to our approach, and we use TF for easy understanding of the proposed concept.}. In general, we use the TF concept as follows: we first compute the number of occurrences\,(i.e., frequency) of each term throughout the whole data set and then choose the highly frequent dimensions. We call this selection method DF\,(document frequency) as in \cite{Tang05}. The reason why we consider TF\,(or DF) in {\sf SSDD-FS} is that, if we select the highly frequent $f$ dimensions, we can obtain relatively small upper bounds $upper(\overrightarrow{U^F},\overrightarrow{V^F})$'s by relatively large $D^2(\overrightarrow{U^F},\overrightarrow{V^F})$'s of Eq.~(\ref{eq:eq2}), and accordingly, we can exploit the filtering effect largely.

As a feature selection using term frequencies, we first consider how frequent each term is in an individual document rather than the whole document set, that is, we first propose the feature selection of exploiting {\it locality\/} of each document. More precisely, for a pair of documents $(\overrightarrow{U},\overrightarrow{V})$, the locality-based selection chooses $f$ dimensions highly frequent in Alice's current vector $\overrightarrow{U}$. This selection is based on the simple intuition that, even without considering whole vectors of the document set, the current vector itself will make a big influence on the upper bound $upper(\overrightarrow{U},\overrightarrow{V})$. In this selection, we can instead use Bob's vector $\overrightarrow{V}$ rather than Alice's vector $\overrightarrow{U}$ as the current vector, or we can also use both Alice's and Bob's vectors $\overrightarrow{U}$ and $\overrightarrow{V}$. Using $\overrightarrow{V}$, however, incurs the additional communication overhead, and thus, in this paper we consider a simple method of using Alice's $\overrightarrow{U}$ as the current vector. We call this selection method {\it LF}(local frequency) since it considers individual (i.e., local) documents rather than whole documents, and we denote the protocol of applying LF to {\sf SSDD-FS} as {\sf SSDD-LF}.

\begin{figure}[hbt]
  \centering
  \psfig{file=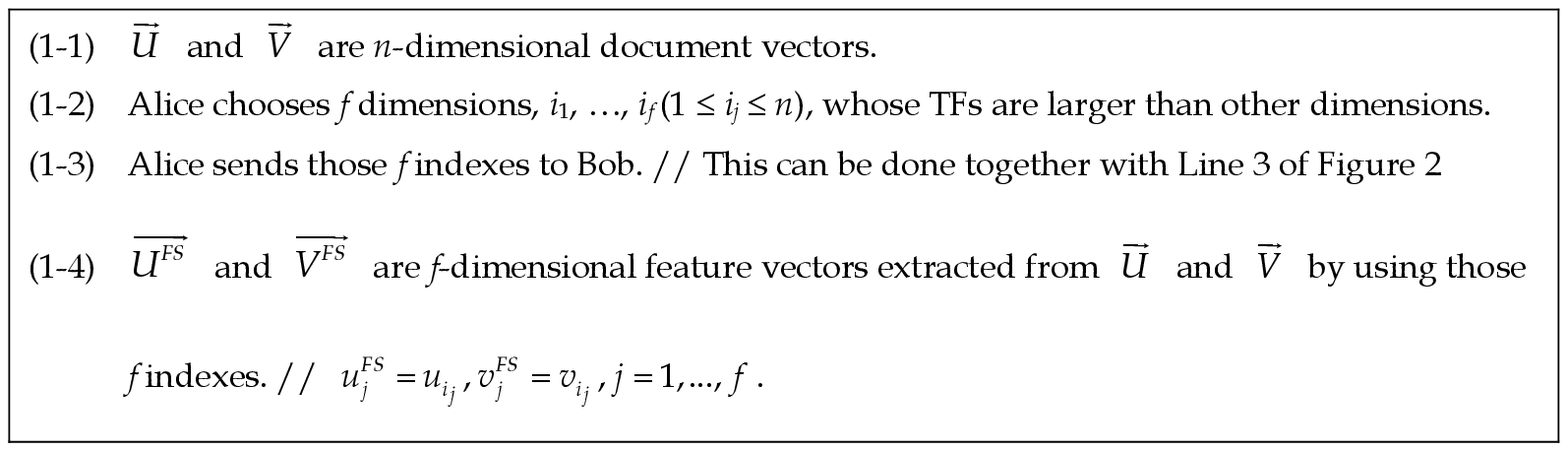,width=6.00in}\\
  \caption{Modification of Line (1) of {\sf SSDD-FS} to implement {\sf SSDD-LF}.}
  \label{fig:fig6}
\end{figure}

{\sf SSDD-LF} exploits the locality by selecting $f$ dimensions for each document at every start time. Figure~\ref{fig:fig6} shows how we implement {\sf SSDD-LF} by modifying Line (1) of {\sf SSDD-FS} of Figure~\ref{fig:fig3}. In Line (1-2), Alice first selects top $f$ frequent dimensions from her current vector $\overrightarrow{U}$. She sends those indexes of the selected $f$ dimensions to Bob in Line (1-3). Thus, they can share the same indexes and obtain $f$-dimensional feature vectors by using the same $f$ indexes in Line (1-4).

We now analyze the computation and communication overhead of feature selection in {\sf SSDD-LF}. As shown in Figure~\ref{fig:fig6}, for each vector $\overrightarrow{U}$, Alice (1) chooses the top $f$ frequent dimensions from $n$ dimensions of $\overrightarrow{U}$ and (2) communicates with Bob to share those $f$ indexes. First, Alice needs the additional computation overhead of $O(n \log f)$  to select top $f$ frequent dimensions from the current $n$-dimensional vector. Second, Alice and Bob need the additional communication overhead to share the $f$ indexes. However, this communication process can be done with Line (3) of {\sf SSDD-Base} of Figure~\ref{fig:fig2}, that is, Alice can send $f$ indexes together with the encrypted vector $\overrightarrow{Z}$ to Bob. The amount of $f$ indexes is much smaller than that of the $n$-dimensional vector, and the overhead of $f$ indexes can be negligible. Thus, we can say that {\sf SSDD-LF} causes the computation overhead of $O(n \log f)$, but the communication overhead can be ignored. In particular, we compare each vector $\overrightarrow{U}$ of Alice with a large number of vectors $\overrightarrow{V}\,(\in \mathbb{V})$ of Bob, and thus, the computation overhead of $O(n \log f)$ can also be ignored as a pre-processing step.

Another considering point in {\sf SSDD-LF} is whether its feature selection process is secure or not. That is, there should be no privacy disclosure when Alice selects $f$ indexes and shares them with Bob. Fortunately, Alice sends only indexes $i_j$ to Bob rather than entry values $u_{i_j}$ of $\overrightarrow{U}$, and the sensitive values $u_{i_j}$ are not disclosed in the selection process. Unfortunately, however, the information that which $f$ dimensions are frequent in $\overrightarrow{U}$ is revealed to Bob. If the user cannot be allowable even this limited disclosure of information, s/he cannot use {\sf SSDD-LF} as a secure protocol. In this case, we recommend to use the previous {\sf SSDD-RP} or the next {\sf SSDD-GF} or {\sf SSDD-HF} as the more secure protocol.

\subsection{GF: Global Frequency} \label{ssec:sec43}
%
{\sf SSDD-LF} of Section~\ref{ssec:sec42} has a problem of considering only Alice's current vector but ignoring all the other vectors of Bob. Due to this problem, {\sf SSDD-LF} exploits the filtering effect for only a part of Bob's vectors, but it does not for most of other vectors. To overcome this problem, in this section we propose another feature selection that uses the whole vector of which each element represents the number of documents containing the corresponding term. Unlike LF of focusing on the current vector only, it considers whole document vectors, and it has characteristics of globality. We call this feature selection {\it GF}\,(global frequency) and denote the GF-based secure protocol as {\sf SSDD-GF}. Actually, GF is the same as DF, which has been widely used as the representative feature selection, and it works as follows. First, let $\overrightarrow{A} = \{a_1, \ldots, a_n\}$ be a whole vector and $a_k$ be a number of documents containing the $k$-th term, that is, $a_k$ be the DF value of the $k$-th term. Then, to reduce the number of dimensions from $n$ to $f$, GF simply selects $f$ dimensions whose DF values are larger than those of the other $(n-f)$ dimensions. We can get the whole vector by scanning all the document vectors once. The traditional DF constructs the whole vector based on the assumption that all the document vectors are maintained in a single computer. In SSDD, however, document vectors are distributed in Alice and Bob, and they do not want to provide their own vectors to each other. Thus, to use GF in SSDD, we first need to present a secure protocol of constructing the whole vector from the document vectors distributively stored in Alice and Bob.

\begin{figure}[hbt]
  \centering
  \psfig{file=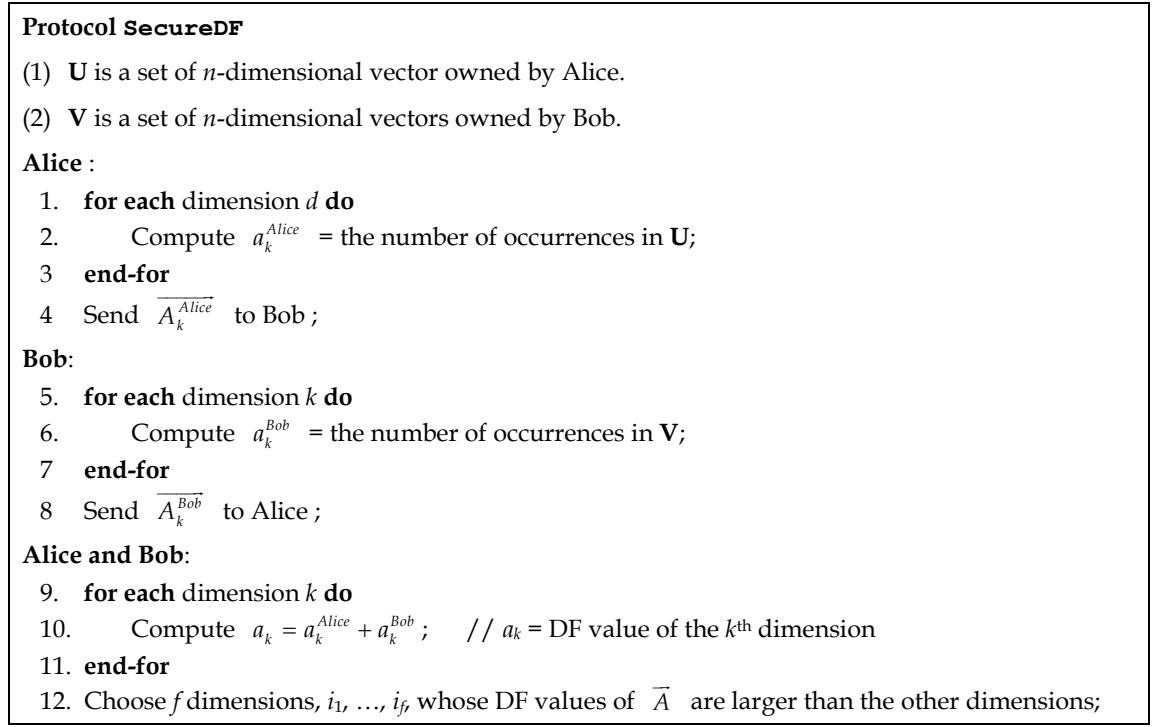,width=6.00in}\\
  \caption{Secure protocol for constructing the whole vector.}
  \label{fig:fig7}
\end{figure}

Figure~\ref{fig:fig7} shows {\bf Protocol} {\sf SecureDF} that securely constructs a whole vector $\overrightarrow{A}$ from Alice's and Bob's document vectors and gets $f$ frequent dimensions from $\overrightarrow{A}$. In Lines 1 to 8, Alice and Bob computes their own whole vectors independently. That is, Alice computes her own whole vector $\overrightarrow{A^{Alice}}$ from her own document set $\mathbb{U}$, and Bob gets $\overrightarrow{A^{Bob}}$ from $\mathbb{V}$. In Lines 4 and 8, they share those whole vectors $\overrightarrow{A^{Alice}}$ and $\overrightarrow{A^{Bob}}$ with each other. In Lines 9 to 11, they then compute the aggregated whole vector $\overrightarrow{A}$ from those vectors. After obtaining the whole vector $\overrightarrow{A}$, Alice and Bob can select $f$ frequent dimensions from $\overrightarrow{A}$. We note that Alice sends $\overrightarrow{A^{Alice}}$ to Bob in Line 4, and Bob sends $\overrightarrow{A^{Bob}}$ to Alice in Line 8. Vectors $\overrightarrow{A^{Alice}}$ and $\overrightarrow{A^{Bob}}$, however, are not exact values of document vectors, but simple statistics, and thus, we can say that {\sf SecureDF} does not reveal any privacy of individual documents. Computation and communication complexities of {\sf SecureDF} are merely $O(n\left|\mathbb{U}\right|+n\left|\mathbb{V}\right|)$ and $O(n)$, respectively. Also, {\sf SecureDF} can be seen as a pre-processing step executed only once for all document vectors of Alice and Bob. Thus, its complexity can be negligible compared with the complexity $(n\left|\mathbb{U}\right|\left|\mathbb{V}\right|)$ of {\sf SSDD-Base}.

\begin{figure}[hbt]
  \centering
  \psfig{file=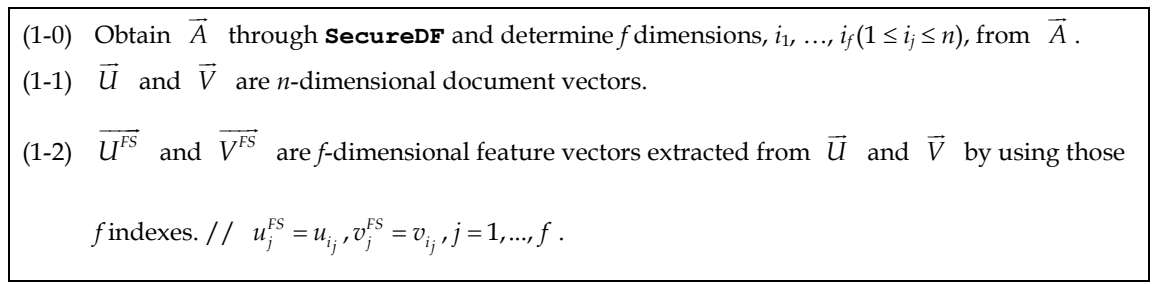,width=6.00in}\\
  \caption{Modification of Line (1) of {\sf SSDD-FS} to implement {\sf SSDD-GF}.}
  \label{fig:fig8}
\end{figure}

We now explain {\sf SSDD-GF} which exploits {\sf SecureDF} as the feature selection. Figure~\ref{fig:fig8} shows how we modify Line (1) of Figure~\ref{fig:fig3} for converting {\sf SSDD-FS} to {\sf SSDD-GF}. In Line (1-0), we first perform {\sf SecureDF} to obtain the whole vector $\overrightarrow{A}$ and determine $f$ indexes which are most frequent in $\overrightarrow{A}$. For current $n$-dimensional vectors $\overrightarrow{U}$ and $\overrightarrow{V}$, Alice and Bob get $f$-dimensional vectors $\overrightarrow{U^{FS}}$ and $\overrightarrow{V^{FS}}$ by using the determined $f$ indexes. As shown in Figure~\ref{fig:fig8}, the current vectors and even their term frequencies are not disclosed to each other, and thus, we can say that {\sf SSDD-GF} is a secure protocol of SSDD.

\subsection{HF: Hybrid Frequency} \label{ssec:sec44}
%
LF and GF proposed in Sections~\ref{ssec:sec42} and \ref{ssec:sec43} have the following characteristics in a viewpoint of the filtering effect. First, LF considers Alice's current vector $\overrightarrow{U}$ only, and thus, the filtering effect will be large for only a part of Bob's vectors whose TF patterns much differ from the current vector, but the effect are less exploited for most of the other vectors. In other words, LF can exploit the better filtering effect than GF when Alice's current vector quite differs from the whole vector in TF patterns. Second, GF considers the whole vector $\overrightarrow{A}$ obtained by {\sf SecureDF} without considering the current vector, and it thus can exploit the filtering effect relatively evenly on many of Bob's document vectors. That is, GF can exploit the better filtering effect than LF when Alice's current vector has the similar characteristics with the whole vector in TF patterns.

To take advantage of both locality of LF and globality of GF, we now propose a hybrid feature selection, called {\it HF}(hybrid frequency). That is, HF uses the current vector for exploiting locality of LF, and at the same time it also use the whole vector for exploiting globality of GF. We then present an advanced secure protocol {\sf SSDD-HF} by applying HF to the {\sf SSDD-FS}. Simply speaking, HF compares current and whole vectors and selects feature dimensions whose differences are larger than those of the other dimensions. In more detail, we select feature dimensions which have one of the following two characteristics: (1) the dimensions which frequently occur in Alice's current vector but seldom occur in the whole vector\,(i.e., whose values are relatively large in the current vector but relatively small in the whole vector); or on the contrary, (2) the dimensions which seldom occur in Alice's current vector but frequently occur  in the whole vector. This is because the larger $|u_i^F - v_i^F|$\,(= the difference between values of the selected feature dimension), the smaller  $upper(\overrightarrow{U^{F}},\overrightarrow{V^{F}})$, i.e., the larger $D^2\!\left(\overrightarrow{U^F},\overrightarrow{V^F}\right)$ of Eq.~(\ref{eq:eq2}), which exploits the larger filtering effect.

However, we cannot directly compare Alice's current vector $\overrightarrow{U}$ and the whole vector $\overrightarrow{A}$ by {\sf SecureDF}. The reason is that $\overrightarrow{U}$ represents ``frequencies of terms'' in a single vector while $\overrightarrow{A}$ represents ``frequencies of documents'' containing those terms. That is, the meaning of frequencies in $\overrightarrow{U}$ differs from that of $\overrightarrow{A}$, and thus, their scales are also different. To resolve this problem, before comparing two vectors $\overrightarrow{U}$ and $\overrightarrow{A}$, we first normalize them using their mean\,($= \mu$) and standard deviation\,($= \sigma$). More precisely, we first normalize $\overrightarrow{U}$ and $\overrightarrow{A}$ to $\overrightarrow{\overline{U}}$ and $\overrightarrow{\overline{A}}$ by Eq.~(\ref{eq:eq9}), and we next obtain the difference vector $\overrightarrow{D} = \{d_1 = |\overline{u_1}-\overline{a_1}|, \ldots, d_n = |\overline{u_n}-\overline{a_n}|\}$. After then, we select the largest $f$ dimensions from $\overrightarrow{D}$ and use them as the features of {\sf SSDD-HF}.

\begin{equation}
\label{eq:eq9}
\overline{u_i} = \frac{u_i - \mu(\overrightarrow{U})}{\sigma(\overrightarrow{U})},
\overline{a_i} = \frac{a_i - \mu(\overrightarrow{A})}{\sigma(\overrightarrow{A})}.
\end{equation}

\begin{figure}[hbt]
  \centering
  \psfig{file=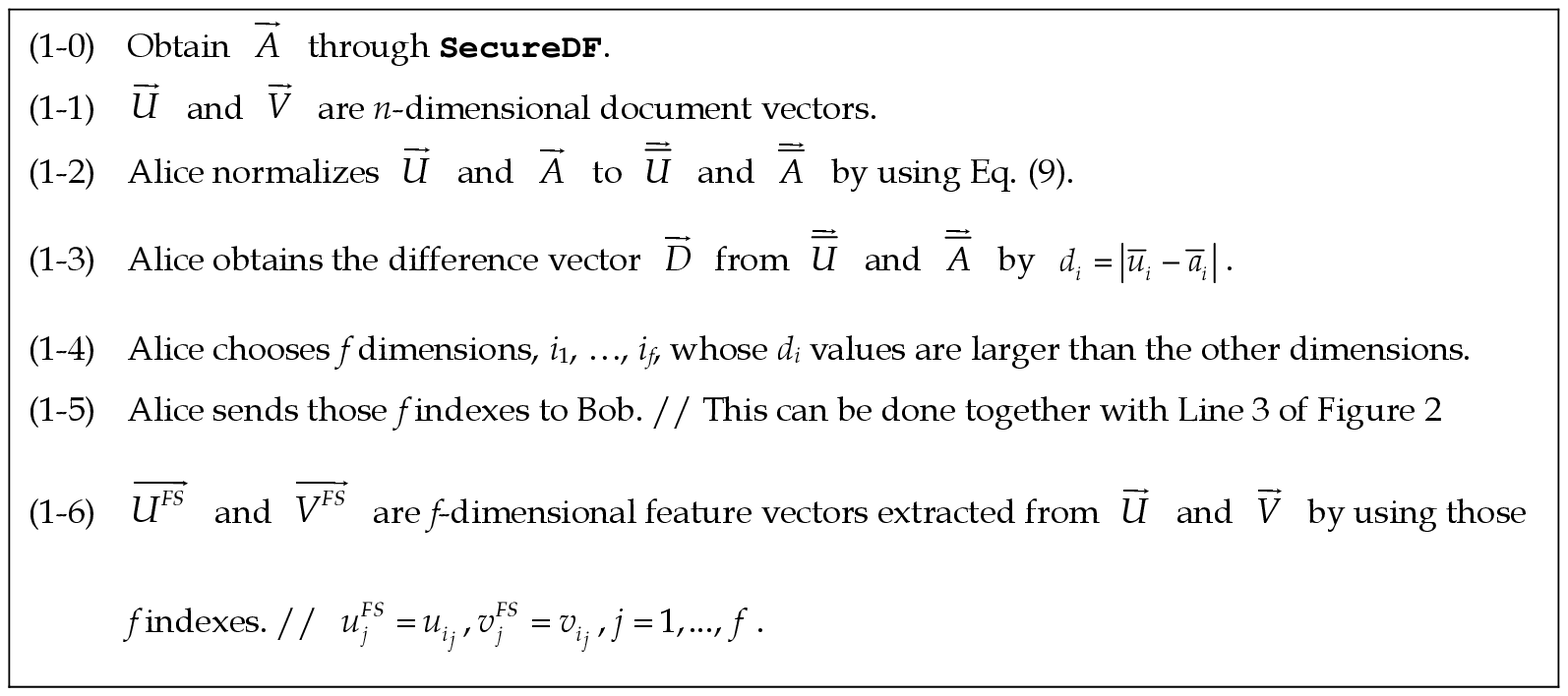,width=6.00in}\\
  \caption{Modification of Line (1) of {\sf SSDD-FS} to implement {\sf SSDD-HF}.}
  \label{fig:fig9}
\end{figure}

Figure~\ref{fig:fig9} shows how we modify Line (1) of {\sf SSDD-FS} in Figure \ref{fig:fig3} to implement {\sf SSDD-HF}. First, as in {\sf SSDD-GF}, Line (1-0) constructs the whole vector $\overrightarrow{A}$ by executing {\sf SecureDF}.
Next, in Lines (1-2) and (1-3), we normalize the current and whole vectors and obtain the difference vector $\overrightarrow{D}$ from those normalized vectors. Finally, in Lines (1-4) to (1-6), Alice chooses $f$ dimensions from the difference vector $\overrightarrow{D}$ and shares those dimensions with Bob. That is, Lines (1-4) to (1-6) are the same as Lines (1-2) to (1-4) of {\sf SSDD-LF} in Figure~\ref{fig:fig6} except that {\sf SSDD-LF} uses the current vector $\overrightarrow{U}$ while {\sf SSDD-HF} uses the difference vector $\overrightarrow{D}$.

The overhead of feature selection in {\sf SSDD-HF} can be seen as the summation of those in {\sf SSDD-LF} and {\sf SSDD-GF}. That is, like {\sf SSDD-GF}, it has the overhead of performing {\sf SecureDF} to obtain the whole vector $\overrightarrow{A}$, and at the same time, like {\sf SSDD-LF}, it has the overhead of choosing the largest $f$ dimensions from the $n$-dimensional difference vector $\overrightarrow{D}$. These overheads, however, can be negligible by the following reasons: (1) as we explained in {\sf SSDD-GF} of Section~\ref{ssec:sec43}, {\sf SecureDF} having $O(n\left|\mathbb{U}\right|+n\left|\mathbb{V}\right|)$ and $O(n)$ of computation and communication complexities can be seen as a pre-processing step executed only once for all document vectors, and its overhead can be negligible in the whole process of SSDD; (2) as we explained in {\sf SSDD-LF} of Section~\ref{ssec:sec42}, the computation complexity $O(n \log f)$ of choosing $f$ dimensions from an $n$-dimensional vector can be ignored since it can also be seen as the pre-processing step. One more notable point is that {\sf SSDD-HF} is a secure protocol like {\sf SSDD-GF} since it uses {\sf SecureDF} and the difference vector which are secure and do not disclose any original values or any sensitive indexes of individual vectors.

\section{Performance Evaluation} \label{sec:sec5}
%
\subsection{Experimental Data and Environment} \label{ssec:sec51}
%
In this section, we empirically evaluate feature selection-based SSDD protocols proposed in Section 4. As the experimental data, we use three datasets obtained from the document sets of UCI repository\,\cite{UCI}. These datasets are KOS blog entries, NIPS full papers, and Enron emails, which have been frequently used in text mining. The first dataset consists of KOS blog entries collected from dailykos.com, and we call it {\it KOS}. KOS consists of 3,430 documents with 6,906 different terms\,(dimensions), and it has total 467,714 terms. The second dataset contains NIPS full papers published in Neural Information Processing Systems Conference, and we call it {\it NIPS}. NIPS consists of 1,500 documents with 12,419 different terms, and it has about 1.9 million terms in total. The third dataset contains e-mail messages of Enron, and we call it {\it EMAILS}. EMAILS consists of 39,861 e-mails with 28,102 different terms, and it has about 6.4 million terms in total.

We experiment five SSDD protocols: {\sf SSDD-Base} as the basic one and four proposed ones of {\sf SSDD-RP}, {\sf SSDD-LF}, {\sf SSDD-GF}, and {\sf SSDD-HF}. In the experiment, we basically measure the elapsed time of executing SSDD for each protocol. In the first experiment, we vary the number of dimensions for a fixed tolerance, where the number of dimensions means $f$, i.e., the number of {\it selected\/} features\,(dimensions) by the feature selection. In the second experiment, we vary the tolerance for a fixed number of dimensions. For these two experiments, we use KOS and NIPS, which have a relatively small number of documents compared with EMAILS. On the other hand, the third experiment is to test scalability of each protocol, and we thus use EMAILS whose number of documents is much larger than those of KOS and NIPS.

The hardware platform is HP ProLiant ML110 G7 workstation equipped with Intel(R) Xeon(R) Quad Core CPU E31220 3.10GHz, 16GB RAM, and 250GB HDD;  its software platform is CentOS 6.5 Linux. We use C language for implementing all the protocols. We perform SSDD in a single machine using a local loop for network communication. The reason why we use the local loop is that we want to intentionally ignore the network speed since different network speeds or environments may largely distort the actual execution time of each protocol. We measure the execution time spent for that Alice sends each document to Bob and identifies its similarity securely. More precisely, we store the whole dataset in Bob and select ten query documents for Alice. After then, we execute each SSDD protocol for those ten query documents and use their sum as the experimental result.

\subsection{Experimental Results} \label{ssec:sec52}
\begin{figure}[hbt]
  \centering
  \psfig{file=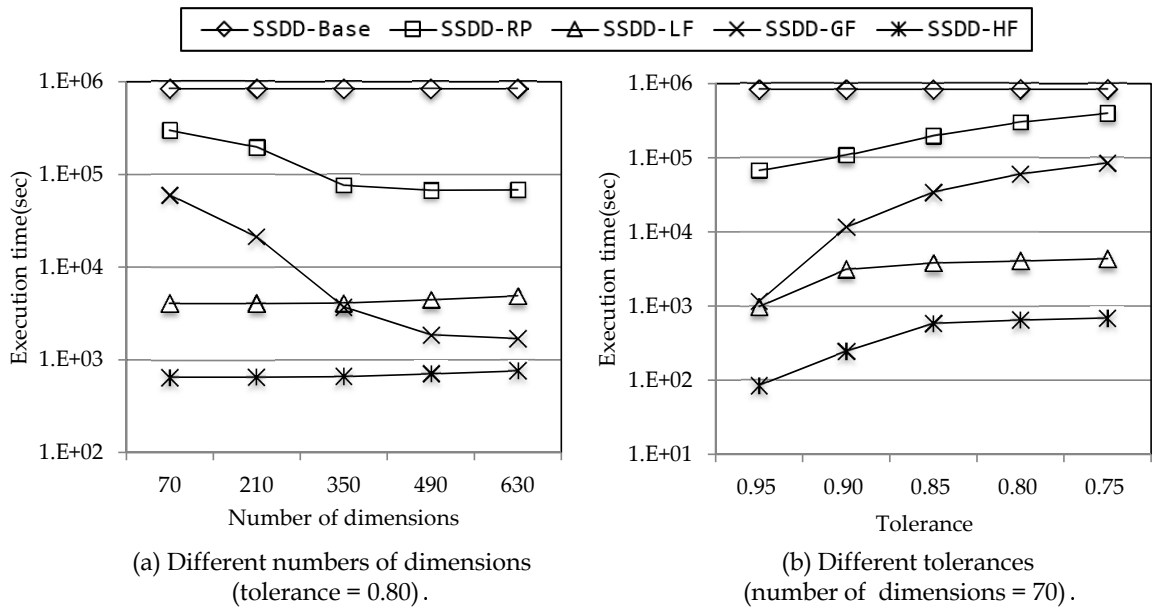,width=6.00in}\\
  \caption{Experimental results for KOS.}
  \label{fig:fig10}
\end{figure}

%
Figure~\ref{fig:fig10} shows the experimental results for KOS. First, in Figure~\ref{fig:fig10}(a), we set the tolerance to 0.80 and vary the number of documents by 70, 210, 350, 490, and 640, which correspond to 1\%, 3\%, 5\%, 7\%, and 9\% of KOS documents. As shown in the figure, $x$ axis shows the number of (selected) dimensions, and $y$ axis does the actual execution time. Note that the $y$ axis is a log scale.

Figure~\ref{fig:fig10}(a) shows that all proposed protocols significantly outperform the basic {\sf SSDD-Base}. Even {\sf SSDD-RP} of selecting features randomly beats {\sf SSDD-Base} by exploiting the filtering effect in the first step of the 2-step protocol. Next, {\sf SSDD-GF} shows the better performance than {\sf SSDD-RP} since it selects the frequently occurred features throughout the whole dataset by using DF. In case of {\sf SSDD-RP} and {\sf SSDD-GF}, we note that, as the number of dimensions increases, the execution time decreases. This is because the more number of dimensions we use, the larger filtering effect we can exploit. {\sf SSDD-LF} of using locality of the current vector also outperforms {\sf SSDD-RP} as well as {\sf SSDD-Base}. In particular, {\sf SSDD-LF} is better than {\sf SSDD-GF} for a small number of dimensions, but it is worse than {\sf SSDD-GF} for a large number of dimensions. This is because only a small number of dimensions make a big influence on the locality of the current vector. Finally, {\sf SSDD-HF} of taking advantage of both {\sf SSDD-LF} and {\sf SSDD-GF} shows the best performance for all dimensions. In Figure~\ref{fig:fig10}(a), we note that the execution time of {\sf SSDD-LF} and {\sf SSDD-HF} slightly increases as the number of dimensions increases. The reason is that, as the number $f$ of dimensions increases, the filtering effect increases relatively slowly, but the overhead of obtaining a current/difference vector and choosing $f$ dimensions from that vector increases relatively quickly.

Second, in Figure~\ref{fig:fig10}(b), we set the number of dimensions to 70\,(1\% of total dimensions) and vary the tolerance from 0.95 to 0.75 by decreasing 0.05. Note that the closer to 1.0 the tolerance is, the stronger similarity we use. As shown in the figure, all proposed protocols significantly improve the performance compared with {\sf SSDD-Base}. In particular, {\sf SSDD-LF} and {\sf SSDD-HF}, which exploits the locality, show the better performance than the other two proposed ones. We here note that, as the tolerance decreases, execution times of all proposed protocols gradually increase. This is because the smaller tolerance we use, the more documents we get as similar ones. That is, as the tolerance decreases, the more documents pass the first step, and thus, the more time is spent in the second step. In summary of Figure~\ref{fig:fig10}, the proposed {\sf SSDD-LF} and {\sf SSDD-HF} significantly outperform {\sf SSDD-Base} by up to 726.6 and 9858 times, respectively.

\begin{figure}[hbt]
  \centering
  \psfig{file=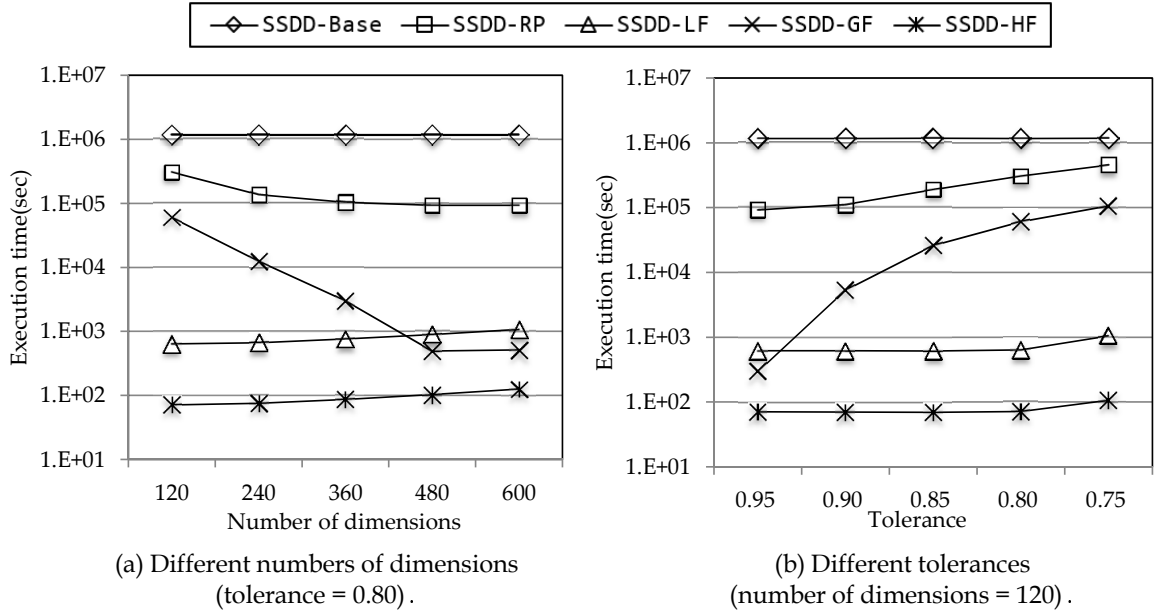,width=6.00in}\\
  \caption{Experimental results for NIPS.}
  \label{fig:fig11}
\end{figure}

Figure~\ref{fig:fig11} shows the experimental results for NIPS. As in Figure~\ref{fig:fig10} of KOS, we measure the execution time of SSDD by varying the number of dimensions and the tolerance. In Figure~\ref{fig:fig11}(a), we set the tolerance 0.80 and increase the number of dimensions from 120\,(1\%) to 600\,(5\%) by 120\,(1\%), where 120 means 1\% of total 12,419 documents. Next, in Figure~\ref{fig:fig11}(b), we set the number of dimensions to 120 and decrease the tolerance from 0.95 to 0.75 by 0.05. The experimental results of Figures~\ref{fig:fig11}(a) and \ref{fig:fig11}(b) show a very similar trend with those of Figures~\ref{fig:fig10}(a) and \ref{fig:fig10}(b). That is, all proposed protocols significantly outperform {\sf SSDD-Base}, and {\sf SSDD-HF} shows the best performance. In Figure~\ref{fig:fig11}, {\sf SSDD-HF} extremely improves the performance compared with {\sf SSDD-Base} by up to 16620 times.

\begin{figure}[hbt]
  \centering
  \psfig{file=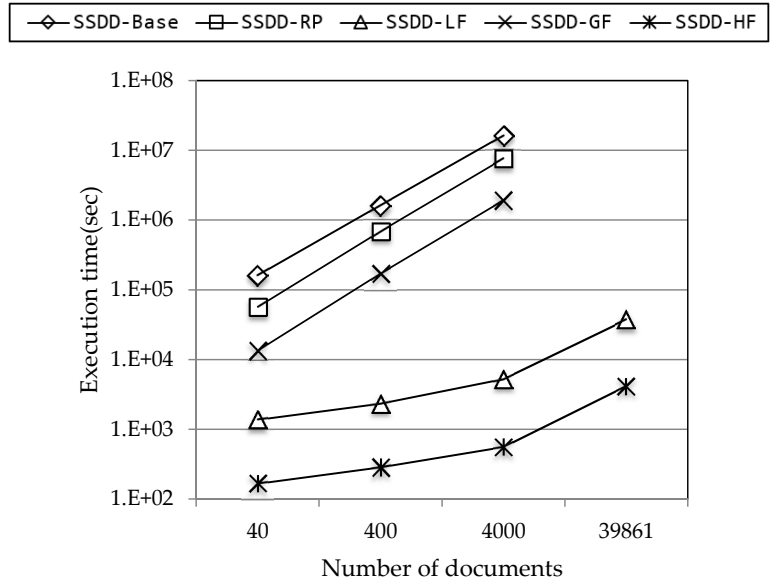,width=4.00in}\\
  \caption{Experimental results of scalability test for EMAILS.}
  \label{fig:fig12}
\end{figure}

Figure \ref{fig:fig12} shows the results for scalability test using a large volume of high dimensional dataset, EMAILS. We set the tolerance and the number of dimensions to 0.80 and 70, respectively, and we increase the number of documents\,(emails) from 40\,(0.1\%) to 39,861\,(100\%) by 10 times. In this experiment, we exclude the results of {\sf SSDD-Base}, {\sf SSDD-RP}, and {\sf SSDD-GF} for the case of 39,861 documents due to excessive execution time. As shown in the figure, like the results of KOS and NIPS, our feature selection-based protocols outperform {\sf SSDD-Base} at all cases, and in particular, {\sf SSDD-LF} and {\sf SSDD-HF} show the best performance regardless of the number of documents. We also note that all proposed protocols show a pseudo linear trend on the number of documents. (Please note that $x$- and $y$-axis are all log scales.) That is, the protocols are pseudo linear solutions on the number of documents, and we can say that they are excellent in scalability as well as performance.

%
\section{Conclusions} \label{sec:sec6}
%
In this paper, we addressed an efficient method of significantly reducing computation and communication overhead in secure similar document detection. Contributions of the paper can be summarized as follows. First, we thoroughly analyzed the previous 1-step protocol and pointed out that it incurred serious performance overhead for high dimensional document vectors. Second, to alleviate the overhead, we presented the feature selection-based 2-step protocol and formally proved its correctness. Third, to improve the filtering efficiency of the 2-step protocol, we proposed four feature selections: (1) RP of selecting features randomly, (2) LF of exploiting locality of a current vector, (3) GF of exploiting globality of all document vectors, and (4) HF of considering both locality and globality. Fourth, for each feature selection, we presented its formal protocol and analyzed its secureness and overhead. Fifth, through experiments on three real datasets, we showed that all proposed protocols significantly outperformed the base protocol, and in particular, the HF-based secure protocol improved performance by up to three or four orders of magnitude. As the future work, we will consider two issues: (1) use of feature extraction\,(feature creation) instead of feature selection for dimensionality reduction and (2) use of homomorphic encryption rather than random matrix for the secure scalar product.

%


\end{document}